\documentclass[preprintnumbers,amsmath,amssymb,prd]{revtex4}
\usepackage{graphicx}
\begin{document}
\preprint{}
\title{Thermal Decay of the Cosmological Constant into Black Holes}
\author{Andr\'es Gomberoff}
\affiliation{Centro de Estudios Cient\'{\i}ficos (CECS), Valdivia, Chile.}
\author{Marc Henneaux}
\affiliation{Physique Th\'eorique et Math\'ematique, Universit\'e
Libre de Bruxelles, Campus Plaine C.P. 231, B--1050 Bruxelles,
Belgium, and \\Centro de Estudios Cient\'{\i}ficos (CECS),
Valdivia, Chile.}
\author{Claudio Teitelboim}
\affiliation{{Centro de Estudios Cient\'{\i}ficos (CECS), Valdivia, Chile.}}
\author{Frank Wilczek}
\affiliation{Center for Theoretical Physics, Massachusetts Institute of
Technology,
Cambridge, MA 02139-4307, USA.\\Centro de Estudios Cient\'{\i}ficos (CECS),
Valdivia, Chile.}
\begin{abstract}
We show that the cosmological constant may be reduced by thermal
production of membranes by the cosmological horizon, analogous to
a particle ``going over the top of the potential barrier", rather
than tunneling through it.
The membranes are endowed with charge associated with the gauge
invariance of an antisymmetric gauge potential.
In this new process, the membrane collapses into a black hole,
thus the net effect is to produce black holes out of the vacuum
energy associated with the cosmological constant. We study here
the corresponding Euclidean configurations (``thermalons"), and
calculate the probability for the process in the leading
semiclassical approximation.

\end{abstract}

\maketitle

\section{Introduction}

One of the outstanding open problems of theoretical physics is to
reconcile the very small observational bound on the cosmological
constant, $\Lambda$, with the very large values that standard high
energy physics theory predicts for it \cite{cosmoproblem}. This
challenge has led to consider the cosmological constant as a
dynamical variable, whose evolution is governed by equations of
motion. In that context one has looked for mechanisms that would
enable $\Lambda$ to relax from a large initial value to a small
one during the course of the evolution of the universe. The
simplest context in which this idea may be analyzed, is through the
introduction of an antisymmetric gauge potential, a three--form \cite{3form}
in four spacetime dimensions. The three--form
couples to the gravitational field with a term proportional to the
square of the field strength. In the absence of sources, the field
strength is constant in space and time and provides a contribution
to the cosmological term, which then becomes a constant of the
motion rather than a universal constant.

Changes in the cosmological constant occur when one brings in
sources for the three--form potential. These sources are
two--dimensional membranes which sweep a three dimensional history
during their evolution (``domain walls"). The membranes carry charge
associated with the gauge invariance of the three--form field, and
they divide spacetime into two regions with different values of
the cosmological term.

The membranes may be produced spontaneously in two physically
different ways. One way is by tunneling trough a potential barrier
as it happens in pair production in two--dimensional spacetime.
The other, is by a thermal excitation of the vacuum analogous to
going ``over the top" of the potential barrier rather than
tunneling through it.

The tunneling process was originally studied in \cite{BT}, and was
further explored in \cite{all}. The purpose of this article is to
study the spontaneous decay of the cosmological constant through the other
 process, namely, the production of membranes due to the
thermal effects of the cosmological horizon.

\begin{figure}[h]
\centering
\includegraphics[width=10cm]{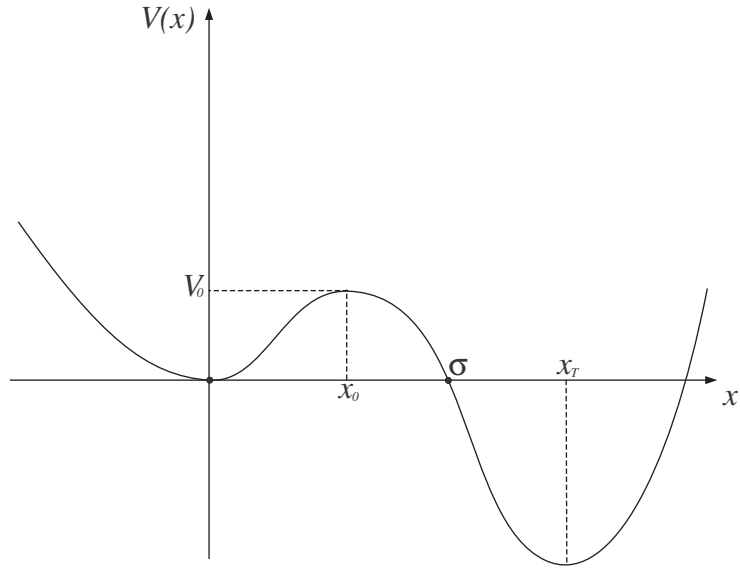}
\caption{The figure shows a one-dimensional potential barrier.
If the particle is initially at the minimum $x=0$ it may end up in the other
side of the barrier,  reaching the lower minimum  $x=x_T$
by two different mechanisms. It can either (i) tunnel through the potential quantum
mechanically or (ii) it can jump over it by a thermal fluctuation when
the barrier is in a thermal environment.}
\label{potential}
\end{figure}

It is useful to visualize the decay process in terms of  its simplest
context, which is a particle in a one dimensional potential
barrier, as recalled in Fig. \ref{potential}. When the barrier is
in a thermal environment, the particle can go from one side of the
barrier to the other by ``climbing over the top" rather than
tunneling.  There is  a probability given by the Boltzmann factor
$e^{-\beta E}$ for the particle to be in a state of energy $E$. If
$E$ is greater than the height of the barrier the particle will
move from one side to the other even classically. The effect is
optimized when the energy is just enough for the particle to be at
the top of the barrier and roll down to the other side. In this
case, the Boltzmann factor is as large as possible while still
allowing for the process without quantum mechanical tunneling. It
turns out that,  in the leading approximation,  the probability
is given by the exponential of the
Euclidean action evaluated on  an 
appropriate classical solution, just as for tunneling
\cite{linde}. In the case of tunneling, the classical solution is
called an instanton and it is time dependent \cite{Coleman}. In
the present case, the classical solution  corresponds to the
configuration in which the particle sits at the top of the
barrier, and thus it is time independent. Since the solution is
unstable, when slightly perturbed the particle will fall half of
the time to the left side and half of the time to the right side.

In more complex situations, ``sitting at the top of the barrier"
is replaced by a ``time independent classical solution with one
instability mode"\cite{linde}.  In the context of gauge theories
such solutions appear in the analysis of violation of
baryon--number conservation and have been called
``sphalerons"\cite{Manton}. In the present case we will use the
name ``thermalon", to emphasize that the static solutions will be
intimately connected with the intrinsic thermal properties of
event horizons in gravitational theory.  We show below that there exists a
thermalon which reduces the cosmological constant through
production of membranes in de Sitter space due to the thermal
properties of the cosmological horizon. In this
thermalon, a membrane  is emitted by the cosmological horizon and
collapses forming a black hole.

\section{The Thermalon}

\subsection{Basic Geometry and Matching Equations}

Once produced, the membrane divides space into two regions with
the interior having a lower value of $\Lambda$. Subsequently, the
membrane evolves with the region of lower $\Lambda$ filling more
and more of the space, thus lowering the value of the cosmological
constant.

In ref. \cite{GHT} we gave the equations of motions of a charged
membrane of tension $\mu$ and charge $q$ coupled to the
gravitational field, and employed them to analyze instantons
associated with tunneling. The same equations will be used here
for the study of thermalons. We reproduce them verbatim here,
together with the corresponding explanations, to make the
discussion self--contained.

We consider an Euclidean spacetime element of the form
\begin{equation}
ds^2= f^2(r) dt^2 + f^{-2} dr^2 + r^2 \left( d\theta^2+\sin^2 \theta d\phi^2 \right) \ .
\label{line}
\end{equation}
The antisymmetric field strength tensor takes the form
\begin{equation}
F_{\mu\nu\lambda\rho}=(dA)_{\mu\nu\lambda\rho}=E\sqrt{g}\epsilon_{\mu\nu\lambda\rho} \ .
\label{}
\end{equation}
The history of the  membrane will divide the spacetime in two
regions, one which will be called the interior, labeled by the
suffix ``-'' and the other  the exterior, labeled by the suffix
``+''. The exterior is the initial region and defines the
``background", while the interior is the final region.  The
boundary  may be described by the parametric equations
\begin{equation}
r=R(\tau) \ \ \ \ \ \ \ t_{\pm}=T_{\pm}(\tau) \ ,
\label{param}
\end{equation}
where $\tau$ is the proper length in the $r-t$ sector, so that its line element reads
\begin{equation}
ds^2= d\tau^2 + R^2(\tau)\left(d\theta^2+\sin^2 \theta d\phi^2 \right) \ ,
\label{memmetric}
\end{equation}
with
\begin{equation}
1=f_{\pm}^2\left(R(\tau)\right) \dot{T}_{\pm}^2 + f_{\pm}^{-2}\left(R(\tau)\right)\dot{R}^2 \ .
\label{propercond}
\end{equation}

In the ``+'' and ``-'' regions the solution of the field equations
read
\begin{eqnarray}
f_{\pm}^2 &=& 1-\frac{2M_{\pm}}{r} - \frac{r^2}{l_{\pm}^2} \ ,  \label{Ds} \\
E_{\pm}^2 &=& \frac{1}{4\pi}\left( \frac{3}{l_{\pm}^2}-\lambda \right) \label{E}\ .
\end{eqnarray}
The actual cosmological constant $\Lambda=3/l^2$ is thus obtained
by adding  $\lambda$, normally taken to be negative, coming from
``the rest of physics" and not subject to change, and the
contribution $4\pi E^2$, which is subject to dynamical equations.
The discontinuities in the  functions $f^2$ and $E$   across the
membrane  are given by
\begin{eqnarray}
f_-^2\dot{T}_- -   f_+^2\dot{T}_+ &=& \mu R  \ .  \label{deltaf}\\
E_{+}- E_{-} &=& q \label{deltaE} \ .
\label{disc}
\end{eqnarray}

Here $\mu$ and $q$ are the tension and charge on the membrane
respectively. Eq.
(\ref{deltaE}) follows from integrating the Gauss law for the
antisymmetric tensor across the membrane, whereas Eq.
(\ref{deltaf}) represents the discontinuity in the extrinsic
curvature of the membrane when it is regarded as embedded in
either the ``-" or the ``+" spaces \cite{Israel}.  In writing
these equations, the following orientation conventions have
been adopted, and
will be maintained from here on: $(i)$ The coordinate $t$
increases anticlockwise around the cosmological horizon, $(ii)$
the variable $\tau$ increases when the curve is traveled along
leaving the interior on its right side.

Equation (\ref{deltaf}) may be thought of as the first
integral of the equation of motion for the membrane, which is thus
obtained by differentiating it with respect to $\tau$ (``equations
of motion from field equations''). Hence, satisfying
(\ref{Ds}-\ref{deltaE}) amounts to solving all the equations of
motion and, therefore, finding an extremum of the action.
More explicitly, the first integral of the equation of motion for the
membrane may be written as,
\begin{equation}
\Delta M = \frac{1}{2}(\alpha^2-\mu^2)R^3 -  \mu f_+^2 \dot{T}_+
R^2 = \frac{1}{2}(\alpha^2+\mu^2)R^3 -  \mu f_-^2\dot{T}_- R^2 \
, \label{masst}
\end{equation}
where  $\Delta M \equiv M_- - M_+$
is the mass difference between the initial and final
geometries and,
\begin{equation}
\alpha^2 = \frac{1}{l_+^2}-\frac{1}{l_-^2}\ . \label{a2}
\end{equation}

 The Euclidean evolution of the membrane
lies between two turning points. Once the initial
mass $M_+$ and the initial cosmological constant
$\Lambda_+=3/l_+^2$ are given, the  turning points, $R$,  are
determined  through,
\begin{equation}
\Delta M = \frac{1}{2}(\alpha^2-\mu^2)R^3 - \varepsilon_+ \mu f_+
R^2 = \frac{1}{2}(\alpha^2+\mu^2)R^3 - \varepsilon_- \mu f_- R^2 \
, \label{mass}
\end{equation}
by setting $\dot{R}=0$ in (\ref{masst}). Here 
we have defined $\varepsilon_{\pm}=\mbox{sgn}\dot{T}_{\pm}$.

\begin{figure}[h]
\centering
\includegraphics[width=8cm]{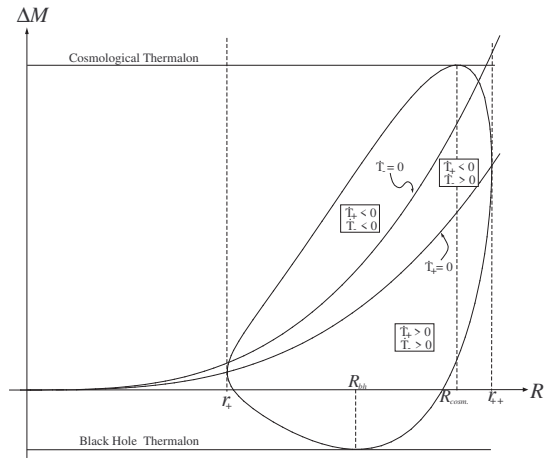}
\caption{The closed curve shows the points $R$ where $\dot{R}=0$
for a given mass gap $\Delta M$. The thermalons are located at the
maximum (cosmological thermalon) and minimum (black hole
thermalon) of the curve. The curves where $R$ is such that for a given $\Delta M$,
$\dot{T}_{\pm}=0$ are also
shown. All the ``+" parameters are held fixed.} \label{masspic}
\end{figure}

The graph of the function (\ref{mass}) for fixed $M_+$, shown in
Fig. \ref{masspic}, will be referred to as the ``mass diagram" for
the decay of Schwarzschild--de Sitter space. There are two
branches which merge smoothly, corresponding to taking both signs
in Eq. (\ref{mass}), much as  $x=\pm \sqrt{1-y^2}$ gives the
smooth circle $x^2+y^2=1$. The lower and upper branches merge at
the intersections with the curve $\Delta M= (1/2)(\alpha^2 - \mu^2)
R^3$ which determines the sign of $\dot{T}_+$, which is
positive below the curve and negative above it. This curve, the
``$\, \dot{T}_+=0$ curve", plays an important role because it
determines which side of the $``+"$ geometry must be retained:
According to the conventions established above,  if
$\dot{T}_+$ is positive, one must retain the side of the membrane
history with (locally) greater values of the radial coordinate; if
$\dot{T}_+$ is negative, one must keep the other side. The curve
$\dot{T}_+=0$ is also shown on the mass diagram, as is the curve
$\dot{T}_- = 0$, which, as can be seen from (\ref{masst}),  has for equation
$\Delta M = (1/2)(\alpha^2 + \mu^2) R^3 $. A similar rule applies 
for determining which side of the minus geometry is retained: it 
is the side of increasing $r$
if $\dot{T}_{-}$ is negative and the other side if it is positive.

\subsection{Black Hole and Cosmological Thermalons}
There are four distinguished points in the mass diagram, two for
which the tangent is vertical and two for which it is
horizontal. The former, located at $r_+$ and $r_{++}$, correspond
to membrane creation through tunneling and they give rise to the
instantons discussed in \cite{GHT}. The latter are thermalons. The
thermalon at the top of the diagram will be called ``cosmological
thermalon" because it is associated with the cosmological event
horizon. The thermalon at the bottom of the mass diagram will be
called ``black hole thermalon".  It needs an initial black hole to
provide the thermal environment.

\begin{figure}[h]
\centering
\includegraphics[width=9cm]{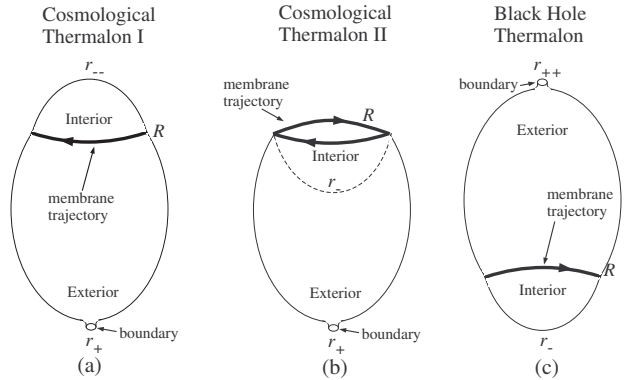}
\caption{Thermalon Geometry. Figures (a) and (b) represent the geometry
of the cosmological thermalon below and above the Nariai threshold respectively,
while (c) depicts the black hole thermalon. Only the $r-t$ section
is shown. In each figure, the radial coordinate increases if one moves upwards.}
\label{euclid}
\end{figure}

For the particular values of $\Delta M$ corresponding to the
thermalons, the two turning points coalesce and the membrane
trajectory is a circle. The geometry of the thermalons is depicted
in Fig. \ref{euclid}.

For the cosmological thermalon, the sign of $\dot{T}_-$ may be either
positive or negative, depending on the values of the parameters.  If $\dot{T}_-$
is negative, one must glue the region of the minus geometry that
contains the cosmological horizon $r_{--}$ to the original
background geometry.
The thermalon has then one black hole horizon
($r_+$) and one cosmological horizon ($r_{--}$).  If, on the other
hand, $\dot{T}_-$ is positive, one must glue the other side of the
membrane history to the original background geometry.  This
produces a solution with two black hole horizons, one at $r_+$ and
one at $r_-$. The inversion of the sign of $\dot{T}_-$ happens when $r_-=r_{--}$,
that is, when the ``-"  geometry becomes the Nariai geometry.
Therefore we will call this particular case
the ``Nariai threshold" and will discuss it in Section IV below.
Note that $\dot{T}_+$ is always negative, hence it
is always the $r_+$-side of the plus  geometry that must be kept.

For the  black hole thermalon $\dot{T}_-$ is always positive.
This is because $\dot{T}_-$ is greater than $\dot{T}_+$,
which is positive in this case.

For both thermalons, the Minkowskian solution is unstable and the
Euclidean solution is stable.  When slightly perturbed, the
Euclidean solution oscillates around the thermalon.
For the Minkowskian case, the membrane can evolve in
two ways: it can start accelerating (rolling down) either towards
the exterior or the interior. If the acceleration is directed
toward the exterior, spacetime will
become filled with the interior (``-") geometry, and the cosmological
term will decrease.  If, on the other hand, it accelerates toward
the interior, then, and as time increases, the exterior will
become the whole geometry and the cosmological constant will
return to its original value. In this case the process will not
have changed anything.

The minimum
and the maximum of the mass curve are Euclidean stable points,
because both correspond
to maxima of the potential on the sense of Fig. 1.   The
cosmological` thermalon corresponds  to  a maximum of the potential
ultimately  because,  as  established  in \cite{CTf,GT},  the internal
 energy of the cosmological horizon is $-M$ rather
than $M$.  Thus, in  the analogy with the potential problem one
should plot $-\Delta M$ in the vertical axis,
so that the potential in Fig. 2 is upside down and the analogy holds.
The instability of both thermalons is proven explicitly
in Appendix \ref{apb}.

Lastly, there is another important distinction between black hole and
cosmological thermalons which resides in how they behave when
gravity is decoupled, that is, when Newton's constant $G$ is taken to vanish.
In that case the black hole thermalon becomes the standard nucleation of
a bubble of a stable phase within a metastable medium, whereas
the cosmological thermalon no longer exists. The decoupling of gravity
is discussed in Appendix \ref{apc}.

\section{Lorentzian continuation}

The prescription for obtaining the Lorentzian signature solution,
which describes the decay process in actual spacetime, is to find
a surface of time symmetry in the Euclidean signature solution and
evolve the Cauchy data on that surface in Lorentzian time. The
fact that the surface chosen is one of time symmetry, ensures that
the Lorentzian signature solution will be real.  Another way of
describing the same statement is to say that one matches the
Euclidean and Lorentzian signature solutions on a surface of time
symmetry.

\begin{figure}[h]
\centering
\includegraphics[width=10cm]{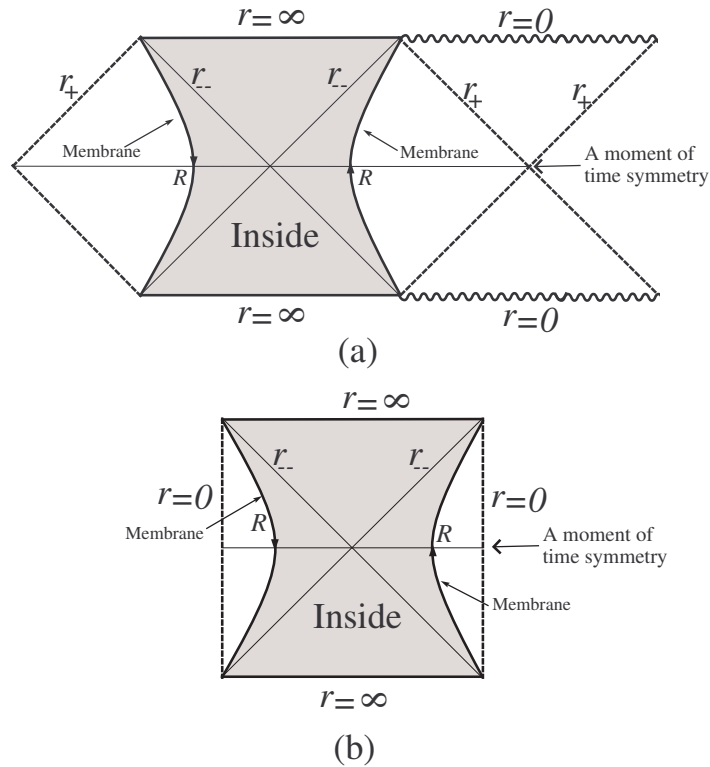}
\caption{Cosmological Thermalon Penrose diagrams.
Figure (b) is the  Penrose diagram for the Lorentzian geometry of the
cosmological thermalon when the initial mass is zero.  The generic case $M_+\neq 0$,
below the Nariai threshold is shown for completeness in figure (a). The doted lines
are the points, $r=r_+$, where the boundary is located in the Euclidean
version of this geometries. Each diagram has two static membranes with opposite
polarities at $r=R$.}
\label{ctdiagram}
\end{figure}

{}For the cosmological thermalon that induces decay of de Sitter
space we will take the surface of time symmetry as the line in
$r-t$ space which, described from the Euclidean side, starts from
$t=t_0$, $r=0$, proceeds increasing $r$, keeping $t=t_0$ until it
reaches the cosmological horizon, and then descends back to $r=0$
along the line $t=t_0+\beta/2$. Thus, the surface of time symmetry
crosses the membrane formation radius, $R$, twice, which implies
that actually {\it two} membranes, of opposite polarities, are formed.  The
cosmological constant is decreased in the finite--volume region
between them. The  Penrose diagram for the Lorentzian
section, below the Nariai threshold, is given in Fig. \ref{ctdiagram}.
To obtain the diagram above the Nariai threshold one should
replace, in Fig. \ref{ctdiagram}, $r_{--}$ by $r_-$ and $r=\infty$
by the black hole singularity $r=0$ of the ``-" geometry.

\section{Action and Probability}

We will be interested in the probability $\Gamma$ per unit of time and unit of spatial
volume for the production of a thermalon.
In the leading semiclassical approximation, that probability is given by
\begin{equation}
\Gamma= A e^{-B/\hbar}(1+{\cal O}(\hbar)) \ ,
\label{pbb}
\end{equation}
where  $-B=I_{\mbox{\tiny thermalon}}$ is the value of the Euclidean action on
the thermalon solution (in our conventions, the sign of the Euclidean action is such
that, in the semiclassical limit, it corresponds to $-\beta F$, the inverse temperature
times the free energy).
The prefactor $A$ is a slowly varying function with dimensions
of a length to the negative fourth power build out of $l_+$, $\alpha^2$ and $\mu$.
The choice of the action depends on the boundary conditions used, and also
involves ``background subtractions", which ensure
that when the coupling with the membrane is turned off  ($\mu=q=0$),
the probability (\ref{pbb}) is equal to unity, since, in that case, $P$ becomes
the probability for things to remain as they are when nothing
is available to provoke a change.

To address the issue of boundary conditions we refer the readers
to figures  \ref{euclid} (a), (b) and (c). There we have
drawn small circles around the
exterior black hole and cosmological horizons respectively,
to indicate that the corresponding  point is treated as a boundary, as discussed
in \cite{CTf,GT}. The point in question takes in each case the place
of spatial infinity for a black hole in asymptotically flat space.
It represents the ``platform" on which the ``external observer"
sits. On the boundary there is no demand that the equations of
motion should hold, and thus it is permissible for a conical
singularity to appear there, as it indeed happens.

Once a boundary is chosen one sums in the path integral over all
possible configurations elsewhere. As a consequence, no conical
singularity is allowed anywhere else but at the boundary.
Therefore, the gravitational action should include a contribution
equal to one fourth of the area of that horizon which is not at the
boundary.

To be definite, consider the cosmological thermalon. In that
case the boundary is placed at $r_+$, and therefore, one is
including the thermodynamical effects of the cosmological horizon
(which is the reason for the term ``cosmological thermalon"). We then fix
at the boundary the value of $r_+$ itself or, equivalently, the
mass $M_+$, in addition to fixing the cosmological constant
$\Lambda_+$.

With these boundary conditions the total action for the problem is just the
standard ``bulk" Hamiltonian action of the coupled system formed by the
gravitational field, the antisymmetric tensor field and the membrane, with
one fourth of the corresponding  horizon area added\cite{BTZ}. This action
includes the minimal coupling term of the three--form field to the membrane,
by which the canonical momentum of the membrane differs from the purely kinetic
(``mass times the velocity") term, and  whose evaluation  needs a definition of the
potential on the membrane. Such a definition requires a mild  form of
regularization which is dictated by the problem itself, and which we
pass to analyze now.

The minimal coupling term is
\begin{equation}
q\int_{V_3} A \ ,
\label{minimal}
\end{equation}
where  the integral extends over the membrane history $V_3$ and $A$ is evaluated on $V_3$.
The potential $A$ is such that the magnitude $E$ of its exterior derivative jumps by $q$ when
passing from the interior (``-" region) of $V_3$ to the exterior (``+" region)  as stated in
Eq. (\ref{deltaE}). We will impose the following conditions on $A$: ($i$) $A$ must be
regular at that origin (horizon) which is in the interior of $V_3$. More precisely,
$A_-$ should be equal to zero up to a regular gauge transformation, in order for the
integral over a very small loop enclosing the horizon to vanish as the loop
shrinks to a point. This is quite straightforward. The subtlety  comes in with $A_+$,
the value of $A$ in the exterior of $V_3$. The fields in the exterior should be those
corresponding to the solution of the equations of motion that would hold everywhere
if the transition where never to occur. Therefore,  we will demand that:
($ii$) the function $A_+$ will be the same, for the given value of $E_+$,
as when the membrane is absent. This means that $A_+$ should be regular at the horizon
in the absence of the membrane, which implies -- as one may show -- that $A$ is
discontinuous across the membrane.
With the above definition of $A$ one may now rewrite the minimal coupling term
(\ref{minimal}) as an integral over the interior $V_4$
 of the membrane, by means of Stoke's formula. This gives,
\begin{equation}
\int_{V_3}A=qE_{\mbox{\tiny av}} = \left(E_+-E_-\right)\frac{1}{2}\left(E_+ +
E_-\right)=\frac{1}{2}\left(E_+^2 - E_-^2 \right)=
\frac{3}{8\pi}\alpha^2 \ ,
\label{adef}
\end{equation}
with $\alpha^2$ given by (\ref{a2}). The appearance of the average field,
$E_{av}$, may be thought of as coming
from defining  the  integral over $V_4$ as the average of the
integrals obtained when the boundary of the $V_4$ is displaced
infinitesimally towards the interior and exterior of the membrane
worldsheet. This is equivalent to ``thickening" the membrane and
taking the boundary half way inside.

It is interesting to point out that,
in the case of membrane production
in flat space, the definition  of the potential $A$ on the
membrane through the ``thickening"  employed in (\ref{adef})
is equivalent to subtracting the background field action $\int F_+^2$.
Thus, we will assume that when (\ref{adef}) is used there remains only
to subtract the gravitational background action,
which is equal to one fourth of the horizon
area in the absence of the membrane (the background Hamiltonian is zero!).
The difference in the horizon areas with and without membrane may be thought
of the change in the available phase space of horizon states induced by
the creation of the membrane.

To be able to write explicitly the form of $I$, one further notices
that, since on-shell the Hamiltonian constraints hold, the bulk
Hamiltonian action reduces to the `$`p\dot{q}$" term. Furthermore,
for both the gravitational and antisymmetric fields, which are time independent
in the exterior and interior of the membrane, $\dot q$ vanishes, and,
therefore,  only the membrane contribution which contains
both the membrane kinetic term and the minimal coupling terms, remains.

With all these observations taken into account, the action
which appears in the probability of  the cosmological thermalon
becomes,
\begin{equation}
I_{\mbox{\tiny thermalon I}}=\frac{1}{4} \left[A(r_{--})-A(r_{++})\right] -
\frac{\mu}{4\pi} V_3 + \alpha^2\frac{3}{8\pi} V_4  \ ,
\label{probability}
\end{equation}
below the Nariai threshold, and
\begin{equation}
I_{\mbox{\tiny thermalon I}}=\frac{1}{4} \left[A(r_{-})-A(r_{++})\right]
-\frac{\mu}{4\pi} V_3 + \alpha^2\frac{3}{8\pi} V_4  \ .
\label{probability2}
\end{equation}
above it.

For the central case of interest in this paper, namely, the decay of de Sitter space
through the cosmological thermalon, we set the initial mass $M_+$ equal to zero.
Note that the cosmological thermalon lies in the upper branch of the mass curve, while
the tunneling decay investigated in \cite{BT}  lies on the lower branch.
Therefore, the thermalon process discussed here and the tunneling
of \cite{BT} are not to be thought of as happening in the same potential barrier in the sense
of the analogy illustrated by Fig. \ref{potential}.
Yet,  the two probabilities  may  be compared, and
the comparison is of interest.
Since there is no black hole in the initial state, the initial
geometry has the full $O(5)$-symmetry and the nucleation process
can occur anywhere in spacetime.  Hence, the computed probability
is a probability per unit of spacetime volume. Note that the thermalon
solution breaks the symmetry down to $O(3) \times
O(2)$, while the tunneling solution of \cite{BT} breaks it to the larger
symmetry group $O(4)$. Hence it is expected to have
higher probability \cite{Coleman2}, which is indeed confirmed by the
analysis given  below.

After de Sitter decays through the thermalon once, the process may happen again, and
again. If the black hole formed after the decay is small, on may, to a
first approximation, ignore its presence and use the same formula for the probability,
taking the final cosmological radius $l_-$ of the previous step as the $l_+$ of the new step.
It is quite alright to  ignore the presence of the black hole when $r_-$ is small because
the probability of the second black hole to be near the first one
will be very small, since it is a probability  per unit of volume.
The approximation will break down when the Nariai bound is in sight.

\section{Nariai threshold}

For given membrane parameters $\mu, q$,
bare cosmological constant $\lambda$ and initial mass $M_+$,
there is a  value $l_N$ of $l_+$  for which the final
geometry becomes the Nariai solution.
In that case, $R= r_- = r_{--}$, since
there is nowhere else for $R$ to be, and thus  the curve
$\dot{T}_-=0$, which always  crosses the mass curve at a root of $f^2_-$, does it now precisely
at the cosmological thermalon nucleation radius $R$.
If one starts from a small $l_+$ one finds the situation illustrated in Fig. 3a. As  $l_+$ increases,
the size of the black hole present in the ``-" region also increases, and at $l_+=l_N$,
the ``-"  geometry becomes the Nariai solution. If $l_+$ increases  further,
and we will refer to this further increase as ``crossing the Nariai threshold",
$\dot{T}_-$ becomes positive, which means that,
as seen from the ``-" side, the orientation of the membrane is reversed.
This implies that one must glue  the part of the ``-" region which one was discarding
below the Nariai threshold to the ``+" region, thus giving rise to the situation
described in Fig. 3b.

It should be emphasized that crossing  the Nariai threshold is not a violent operation.
The bound
\begin{equation}
M_- \leq \frac{l_-}{3 \sqrt{3}}, \label{Nariai}
\end{equation}
for the existence of a de Sitter black hole is maintained throughout ($f_+(R)$ is real  since $R < l_+$, which
implies, using  (\ref{mass}),  that $f_-(R)$ is also real and hence  the function $f_-^2$ has two positive real roots).
The action, and hence the probability, remains continuous. However, the nature of the
thermalon geometry is now somewhat different, since the ``-" geometry  has a black hole
horizon instead of a cosmological horizon.  As one moves from the boundary $r_+$ towards
the membrane, ther radius $r$ of the attached  $S^2$ increases.  But,  after crossing the membrane,
 it starts decreasing --  in contradistinction to what happens
below the  threshold,  --  until it reaches its minimun
value at the black hole radius of the ``-" geometry.

\section{Decay of de Sitter space through the cosmological thermalon}

\subsection{Nucleation radius and mass of final state black hole}

We now focus on the central case of interest in this paper,
namely, the decay of de Sitter space through the cosmological
thermalon.
The cosmological thermalon radius of nucleation, $R$, may be
obtained by differentiating Eq, (\ref{mass}) in its plus sign
version,
\begin{equation}
\frac{3}{2}(\alpha^2-\mu^2) R + 2\mu f_+ + \mu {f_+}^\prime R =  0
\ \ . \label{cteq}
\end{equation}
When $M_+=0$, Eq. (\ref{cteq})  can be rewritten in terms of the dimensionless auxiliary
variable $x=f_+ l_+/R$, or, equivalently,
\begin{equation}
R^2 = \frac{l_+^2}{1+x^2} \ . \label{newzerom}
\end{equation}
Then, (\ref{cteq}) gives,
\begin{equation}
x=\frac{3}{4}\left[-\gamma + \left(\gamma^2+\frac{8}{9}\right)^{1/2}\right] \ ,
\label{x}
\end{equation}
where,
\begin{equation}
\gamma=\frac{l_{+}(\alpha^2 - \mu^2)}{2\mu} \ .
\label{gamma}
\end{equation}

{}From Eq. (\ref{mass}) we obtain for
the mass of the black hole appearing in the final state,
\begin{equation}
M_- = \frac{\mu l_+^2}{3x}(1+x^2)^{-1/2} \ . \label{Mminus}
\end{equation}

The Nariai threshold radius $l_+=l_N$ may be also evaluated explicitly. It is given by
\begin{equation}
\frac{1}{ l_N^2} = \frac{1}{2}\left(\mu^2 + 8 \pi q^2 + \sqrt{8
\pi q^2 (6 \pi q^2 - 2 \lambda + 3 \mu^2)}\right) \ . \label{alpha1}
\end{equation}

\subsection{Response of the final geometry to changes in the initial cosmological constant}

\begin{figure}[h]
\centering
\includegraphics[width=10cm]{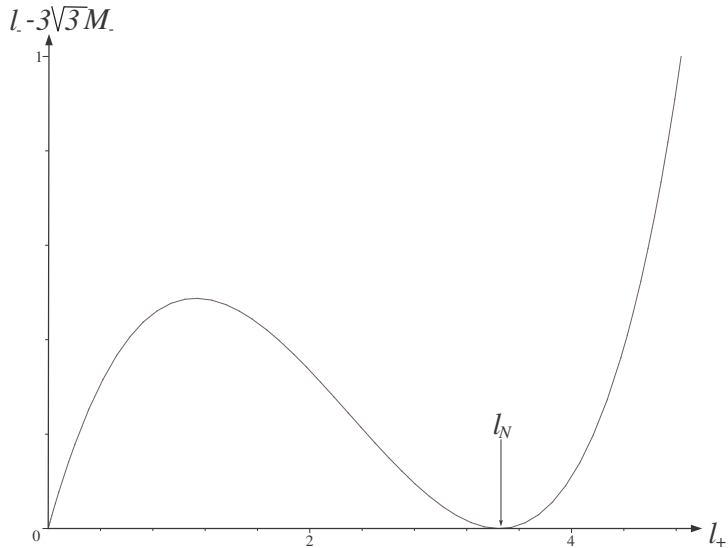}
\caption{The quantity $l_- - 3\sqrt{3}M_-$ is shown in the graph as a function
of the initial cosmological radius $l_+$. It vanishes when the final geometry is that
of Nariai, at $l_+=l_N$, and it goes to infinity when the final geometry has
vanishing cosmological constant, at $l_+=l_\infty$. Here we have set, in Planck units,
$q=0.01$, $\mu=0.3$ and $\lambda=-1$.}
\label{nc}
\end{figure}

{}For $l_+ < l_N$ one easily verifies that, at $R$,
$t_- < \Delta M$ and hence  $\dot{T}_-<0$. Therefore the thermalon
geometry is the one pictured in Fig. \ref{euclid} (a).
As $l_+$ increases the final state approaches the Nariai solution.
We have plotted in Fig. \ref{nc} the quantity $l_{-} - 3
\sqrt{3} M_-$ as a function of $l_+$.  As $l_+$ crosses
the Nariai value $l_N$, the situation becomes the one depicted in Fig.
\ref{euclid} (b). If $l_+$ increases further, the function $l_-$ and thus also
$l_{-} - 3 \sqrt{3} M_-$ blows up for some value $l_\infty > l_N$
of $l_+$; for that value, the final cosmological constant
$l_-$ is infinite and $\Lambda_-$ vanishes .  Above $l_\infty$
the final cosmological constant is negative and thus  a
transition from de Sitter space to Schwarzschild-anti-de Sitter space takes place.
The transition probability is well defined because, since $\dot{T}_-$ is still positive,
a finite-volume piece of Schwarzschild-anti-de Sitter space enters in the action, namely the one
between the membrane and the  black hole horizon $r_-$.

\begin{figure}[h]
\centering
\includegraphics[width=10cm]{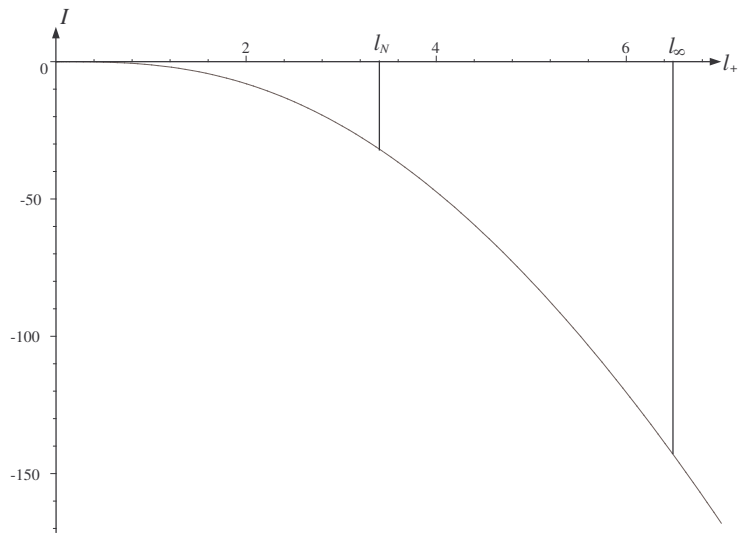}
\caption{ The graph depicts the value of the action for the cosmological thermalon
as a function of the initial cosmological radius. We have used the same values
of the parameters as those used in Fig. \ref{nc}.}
\label{actionfig}
\end{figure}

The radius $l_+=l_\infty$, for which the final cosmological constant vanishes it is given by
 \begin{equation}
  \frac{1}{l_{\infty}^2}= \frac{4}{3}q\left(\pi q + \sqrt{-\pi\lambda}\right) \ .
 \label{linfty}
 \end{equation}
Because the horizons $r_-$, $r_{--}$ and inverse temperature
$\beta_-$ of the final state are determined by algebraic equations
of a high degree, we have found it necessary to compute the action  $I(l_+)$
by direct numerical attack.
The result gives a curve of the form shown in figure
\ref{actionfig}. One sees that the probability
 decreases very quickly as $l_+$ increases.
Since, as the graph
 shows, the action is very small
for large cosmological constant the process is not
exponentially suppressed at the beginning.  As the process goes
on, the action becomes monotonously more negative and
the probability becomes exponentially  suppressed. The action
is continuous both at the Nariai value $l_N$ and at the critical
point $l_\infty$ where the final cosmological constant becomes
negative.

\subsection{Small charge and tension limit}

In order to avoid fine tuning it is necessary to assume
that the jumps in the cosmological constant are of the order of
the currently observed value $\Lambda_{\mbox{\tiny obs}}$, which is, in Planck
units adopted from now on, $\Lambda_{\mbox{\tiny obs}} \sim 10^{-120}$.
Thus we take
$\alpha^2 \sim 10^{-120}$ (``small jump condition"). The bare
cosmological constant itself could be as big as the Planck scale
($\vert \lambda \vert \sim 1$) unless protected by some
broken symmetry, e.g. $\vert \lambda \vert \sim 10^{-60}$
(SUSY).  From (\ref{E}) we find that the $E$ field is of the order of
$\sqrt{-\lambda}$, which gives, using (\ref{adef}), $q \sim
10^{-120}/\sqrt{-\lambda}$:  the small jump condition requires a small
charge $q$.   As the decay of the
cosmological constant proceeds, the radius of the universe $l_+$
goes from ``small" values $\sim (\sqrt{\vert \lambda \vert})^{-1}$
to large values $\sim 10^{60}$ and the dimensionless product
$\alpha^2 l_+^2$ varies from $\sim 10^{-120}$ (Planck) or $\sim
10^{-60}$ (SUSY) to a quantity of order unity.  We will assume that,
\begin{equation}
l_+ ^2 \alpha^2  \ll 1 \label{limit1} \ ,
\end{equation}
which holds during the whole decay of the cosmological constant, beginning
from early stages $l_+ \sim 1$ (Planck), or $l_+\sim 10^{30}$ (SUSY), all the way
up to ``almost" the late stages $l_+ \sim 10^{58}$, say.
We shall call (\ref{limit1}) the ``small charge limit".

We will also assume
\begin{equation}
l_+\mu \ll 1 \ ,
\end{equation}
which we shall call the ``small tension limit".

Within the small charge and tension limits there are two interesting subcases
which are amenable to analytical treatment:

$(a)$ First, we may assume that
\begin{equation}
\frac{l_+  \alpha^2}{\mu} \ll 1 \ ,
\label{first}
\end{equation}
This limit may be achieved, for instance, by taking $\mu$ to
be of the order of $\alpha$, that is, $\mu\sim 10^{-60}$, again, up  late stages
of the evolution of $l_+$.
In this limit,
\begin{equation}
\gamma \ll 1  \ .
\end{equation}
Direct computations yield
\begin{equation}
R^2 \approx \frac{2}{3} l_+^2
\left(1+\frac{\gamma}{\sqrt{2}}\right) \label{appfr} \  , \ \ \
\frac{M_-}{l_+} \approx \frac{2\sqrt{3}}{9} \mu l_+
\left(1+\sqrt{2}\gamma\right) \ .
\end{equation}
The nucleation radius is roughly $\sqrt{2/3} l_+$
and the initial and final universe radii $l_+$ and $l_-$ are equal
to leading order. Since both $\alpha^2 l^2_+$ and $\mu^2 l_+^2$
are small, Eq. (\ref{alpha1}) implies $l_+ \ll l_N$ so that the
final geometry contains the cosmological horizon $r_{--}$. 

{}We then get for the action $I_{\mbox{\tiny Thermalon}}$,
\begin{equation}
-\frac{8\pi\sqrt{3}}{9}l_+^2 (\mu l_+) \ \ , \label{actionnt1}
\end{equation}
which can be compared to the tunneling--process action with the
same parameters
\begin{equation}
-\frac{\pi}{2}l_+^2 (\mu l_+). \label{actionbt1}
\end{equation}
We conclude that in this regime the tunneling--process
is more probable than the cosmological thermalon.

$(b)$ Conversely, we may assume,
namely,
\begin{equation}
\frac{\mu}{l_+  \alpha^2} \ll 1 \ ,
\label{second}
\end{equation}
which is equivalent to,
\begin{equation}
\gamma \gg 1  \ .
\end{equation}
This limit may be achieved, for instance,
 by setting $q\sim\mu$ (``BPS condition")
 for late stages of the evolution of the cosmological
 constant ($l_+>10$ (Planck), $l_+>10^{31}$ (SUSY)).
 Here we get for the nucleation radius and final mass,
\begin{equation}
R^2 \approx l_+^2 \left(1 - \frac{1}{9\gamma^2} \right)
\ \ \ \ \ \ , \frac{M_-}{l_+} \approx \mu l_+ \gamma \approx \alpha^2 l_+^2 \ \ ,
\end{equation}
which shows that, in this limit, the nucleation radius gets close to the cosmological
horizon. The value of the action for this process is, to leading order,
\begin{equation}
 - \frac{4\pi l_+^2}{9}\left(\frac{\mu}{\alpha^2 l_+}\right)^2 \ ,
\label{actionnt2}
\end{equation}
which may be compared with the action for the tunneling process in the same
limit,
\begin{equation}
-4\pi l_+^2
\left(\frac{\mu}{\alpha^2 l_+}\right)^4 (\alpha^2 l_+^2)\ \ .
\label{actionbt2}
\end{equation}
We again see in this regime that the tunneling--process 
is more probable than the
cosmological thermalon.

\section{Can the thermalon account for the small present value 
of the  cosmological term ?}

The cosmological thermalon has a distinct advantage over the instanton 
proposed in \cite{BT} as a mechanism for relaxing the cosmological constant in
that it does not have the so called ``horizon problem". Indeed, since 
for the thermalon the cosmological constant is reduced in the region with bigger values 
of the radial coordinate, and then the membrane proceeds to collapse, one is sure
that, even though the universe is expanding (and even more because it is so!), 
the whole of the universe will relax its cosmological constant. 

The other problem that
the instanton has is that the rate of membrane nucleation
was too small to account for the present small value of the cosmological term.
It is not clear at the moment of this writing whether the thermalon will 
also be an improvement in this regard. Indeed, to make a mild assessment 
of whether the rate is sufficiently strong, we 
first recall that, in Planck units, the observed value of the cosmological constant 
now is $10^{-120}$ and the age of the universe $10^{60}$. Even though 
the process of relaxation of the cosmological constant 
may have not occurred through the entire life span of the post Big-Bang universe, we may use
for very rough estimates $10^{60}$ as the time available for the process to keep occurring. 
If we assume that the cosmological radius started at the Planck scale, 
$l_+=1$ and at present has the value
$10^{60}$, and recall that each bubble nucleation reduces $l_+^{-2}$ by $\alpha^2$, one
needs $10^{120}$ events to occur during $10^{60}$ Planck time units. Furthermore, 
since at any given time the volume of space available 
for the location of the center of the bubble is 
of the order of $l_+^3$ (spatial volume in the comoving frame of de Sitter space),
we conclude that the rate $\Gamma$ is given by
\begin{equation}
\Gamma \sim \frac{10^{120}}{10^{60}l_+^3} =\frac{10^{60}}{l_+^3} \ \ \ \ (\mbox{Planck units})\ .
\label{P2}
\end{equation}
In order for Eq. (\ref{P2}) to be realizable for 
$l_+$ in the range $1<l_+<10^{60}$ we get, using (\ref{pbb}), 
\begin{equation}
10^{-120}< Ae^{-B}<10^{60}   \ \ \ .
\label{k}
\end{equation}
One must keep in mind that as stated above, the parameter $\alpha^2$ is fixed to be 
$10^{-120}$ and that the parameter $B$ should not be too small in order 
for the semiclassical approximation to be valid, which imposes a constraint
relating the other two parameters $\mu$ and $l_+$.  It is therefore necessary
to perform a careful analysis of the prefactor $A$ to see whether
the inequality (\ref{k}) can be satisfied with an 
$l_+$ within the range available in the history 
of the universe. If this is not so, one could not argue that the thermalon
can account for all of the relaxation of the cosmological constant
and, a fortiori, one could not argue that all of the vacuum energy was condensed 
into black holes.

\section{Conclusions}

In this paper, we have exhibited a new process through which the
cosmological constant can decay.  At the same time, a black hole
is created. The classical solution describing the process is an unstable
(``ready-to-fall") static solution which we have called
``cosmological thermalon"; it is an analog of the sphaleron of
gauge theories. Gravity and, in particular,  the thermal 
effects of the cosmological horizon is essential for
the existence of the solution, which disappears in the flat space
limit.  

The net effect of the process is thus to transform
non localized dark energy into localized dark matter, thus providing
a possible link between the small present value of the cosmological constant and the
observed lack of matter in the universe.
Of course, the emergence in this way of (nearly) flat space as a natural 
endpoint of dynamical evolution is most intriguing in view of 
cosmological observations.

\acknowledgments

The main results presented in this paper where reported 
at the School on Quantum Gravity, Valdivia, Chile, 
4-14 January 2002, and also at the Tenth Marcel Grossmann Meeting 
on General Relativity, Rio de Janeiro, Brazil, July 20-26, 2003.
 
After this paper was finished we became aware of the interesting
work of J. Garriga and A. Megevand \cite{garriga}, where ``\ldots a 
`static'instanton, representing pair creation of critical bubbles 
- a process somewhat analogous to thermal activation in flat 
space\ldots" is discussed.

We are grateful to the referee for useful comments that led to
an improvement of the first version of this manuscript.
This work  was funded by an institutional grant to CECS of the
Millennium Science Initiative, Chile, and also benefits from the
generous support to CECS by Empresas CMPC. AG gratefully
acknowledges support from FONDECYT grant 1010449  and from
Fundaci\'on Andes. AG and CT  acknowledge partial support under
FONDECYT grants 1010446 and 7010446.  
The work of MH is partially supported by
the ``Actions de Recherche Concert{\'e}es" of the ``Direction de
la Recherche Scientifique - Communaut{\'e} Fran{\c c}aise de
Belgique", by IISN - Belgium (convention 4.4505.86), by a ``P\^ole
d'Attraction Universitaire" and by the European Commission RTN
programme HPRN-CT-00131, in which he is associated to K. U.
Leuven.

\appendix

\section{Instability of thermalons}

\label{apb}

It is illuminating to see explicitly that both the minimum
and the maximum of the mass curve are Euclidean stable points, since ``unthoughtful" analogy
with a standard potential problem  might have suggested  that only the minimum of the mass
curve should be a stable Euclidean equilibrium.

Consider  first the black hole thermalon and perturb
it, $R_{\mbox{\tiny BHT}} \rightarrow R_{\mbox{\tiny BHT}} + \eta(\tau)$.  The mass
equation yields,
\begin{equation} \Delta M(R_{\mbox{\tiny BHT}}) + \delta m = \frac{1}{2} (\alpha^2 -
\mu^2) R^3 - \mu \sqrt{f_+^2 - \dot{R}^2} \, R^2
\label{AAB}\end{equation}
where $\delta m$ is the mass perturbation. Using that $\Delta M$ is extremum at the thermalon, one finds,
\begin{equation}
\delta m = \frac{1}{2} \left.\frac{\partial^2 \Delta M}{\partial
R^2}\right|_{\mbox{\tiny BHT}} \eta^2 + \left.\frac{\mu R^2}{2
f_+} \right|_{\mbox{\tiny BHT}}\dot{\eta}^2 \label{mequ} \end{equation}
Because the second derivative of $\Delta M$ is positive at the minimum
$R_{\mbox{\tiny BHT}}$, the right-hand side of (\ref{mequ}) is non-negative, so that
the perturbation $\eta(\tau)$ is forced to remain in the bounded range
$\vert \eta \vert \leq \sqrt{2 \delta m(\partial^2 \Delta M/ \partial R^2)^{-1}}$,
which implies stability.

For the cosmological thermalon, it is now the upper branch of the mass curve which is
relevant, and (\ref{AAB}) is replaced by
\begin{equation} \Delta M(R_{\mbox{\tiny CT}}) + \delta m = \frac{1}{2} (\alpha^2 -
\mu^2) R^3 + \mu \sqrt{f_+^2 - \dot{R}^2} \, R^2
\label{BB}\end{equation}
so that, instead of (\ref{mequ}) one has
\begin{equation}
\delta m = \left.\frac{1}{2} \frac{\partial^2 \Delta M}{\partial R^2}
\right|_{\mbox{\tiny CT}} \, \eta^2 - \left.\frac{\mu R^2}{2 f_+}\right|_{\mbox{\tiny CT}}
\dot{\eta}^2 \label{AA} \end{equation}
At the  cosmological thermalon, $\Delta M$ is maximum and so its second derivative is
negative.  Again the perturbation $\eta$ is
bounded and the Euclidean solution is stable.

The signs in (\ref{AA}) are in agreement with the proof given in \cite{CTf,GT},
that the internal energy of the cosmological horizon is $-M$, thus the
``thoughtful" analogy with the potential problem simply
amounts to realize that for the cosmological thermalon one should plot
$-\Delta M$ in the vertical axis, so that the ``potential" in Fig. 2 is
upside down and the analogy holds.

We end this appendix computing the frequency of oscillations around the cosmological
thermalon. From (\ref{AA}) we find that those are given by
$$
\omega^2 =  \left.\left(\frac{\partial^2 \Delta M}{\partial R^2}\right)
\frac{f_+}{\mu R^2}\right|_{\mbox{\tiny CT}} \ ,
$$
which may be evaluated in terms of the nucleations radius $R$ of the cosmological thermalon,
\begin{equation}
\omega^2=\frac{2l_+^2-R^2}{R^2(l_+^2-R^2)} \ .
\label{omega}
\end{equation}
Note that there is a resonance in the limit when  $R$ approaches the 
cosmological radius $l_+$. This  may be achieved
when the BPS condition of Eq. (\ref{second}) is satisfied 
($\mu/l_+\alpha^2<<1$, which holds, for instance, setting $q\sim\mu$). 
In that case we get,
\begin{equation}
\omega=\frac{3}{2}\frac{\alpha^2}{\mu} \ ,
\label{omegabps}
\end{equation}
and, from (\ref{second}) we see that, in fact, $\omega>>l_+^{-1}$.
\section{Gravitation Essential for Existence of Cosmological Thermalons.}
\label{apc}

An interesting feature of the cosmological thermalon is that it
does not exist in the limit of no gravity, where Newton's constant $G$ is taken to
vanish.  Indeed, when $G$ is explicitly written, the mass equation
becomes
\begin{equation}
 G \Delta M = \frac{1}{2} (\alpha^2 - G^2 \mu^2)R^3 \pm G \mu f_+ R^2
\end{equation}
with
\begin{equation}
\alpha^2 = \frac{4 \pi G}{3} (E^2_+ - E^2_-)
\end{equation}
and
\begin{equation}
f_+^2 = 1 -\frac{2GM_+}{R} - \frac{R^2}{l_+^2}\, , \; \; \; \; \;
\frac{3}{l_+^2} =  \lambda + 4 \pi G E^2_+
\end{equation}
Note that the bare cosmological constant $\lambda$  depends on $G$ through
\begin{equation}
\lambda=8\pi G \rho_{vac} \ ,
\label{bare}
\end{equation}
where $\rho_{vac}$ is the vacuum energy density coming from ``the rest of physics".
Therefore, in the limit $G \rightarrow 0$, the lower
branch of the mass curve becomes,
\begin{equation}
 \Delta M  = -V(R) = -  (\mu R^2 - \nu R^3) \, , \; \; \; \; \;  \nu = \frac{2
\pi}{3} (E^2_+ - E^2_-) >0
\label{V}
\end{equation}
The (Minkowskian) potential $V(R)$ appearing
in (\ref{V}) exhibits the competition between surface and volume
effects which gives rise to the nucleation of a bubble of a stable phase
within a metastable medium.
 Thus, in the no-gravity limit, the distinguished points
 in the lower branch of the mass curve have a clear interpretation:
 the black hole thermalon is the unstable static solution sitting
 at the maximum of the potential, while the instanton becomes the standard
 bounce solution for metastable vacuum decay.

 On the other hand, the upper branch of the mass curve yields
\begin{equation}
V(R) = \mu R^2 + \nu R^3 \ ,
\end{equation}
and has no maximum.  More precisely, the maximum, $R$, has gone
to infinity ($R \sim 1/\sqrt{G}$).  There is therefore no
zero-gravity limit of the cosmological thermalon.  One may
understand the behavior of the potential by enclosing the
bubble in a sphere of radius $L$ and recalling that in this case,
the change from the metastable to the stable phase occurs
in the {\it exterior} of the bubble (i.e., on the side with bigger
values of the radial coordinates).
This yields the potential $V(R) = \mu R^2 - \nu (L^3 - R^3)$
which has the correct volume dependence for bubble nucleation.
The extra term  $- \nu L^3$ is constant, and infinite in the
limit $L \rightarrow \infty$.
Gravity changes the shape of  the potential and makes it finite, so that
there is an unstable static solution, the cosmological thermalon.

\end{document}